\documentclass[10pt]{article}
\usepackage{amssymb,latexsym}
\usepackage{amsmath,amsbsy}
\usepackage{epsfig}

\DeclareMathOperator{\pperp}{\mathbf{p}^{\perp}}
\DeclareMathOperator{\pperpone}{\mathbf{p}^{\perp}_{1}}
\DeclareMathOperator{\pperptwo}{\mathbf{p}^{\perp}_{2}}
\DeclareMathOperator{\Pperp}{\mathbf{P}^{\perp}}

\DeclareMathOperator{\Pplus}{P^{+}}
\DeclareMathOperator{\pplus}{p^{+}}

\DeclareMathOperator{\kperp}{\mathbf{k}^{\perp}}
\DeclareMathOperator{\kplus}{k^{+}}
\DeclareMathOperator{\kaperp}{\mathbf{\kappa}^{\perp}}
\DeclareMathOperator{\kaplus}{\kappa^{+}}
\DeclareMathOperator{\Pee}{P^{\mu}}
\DeclareMathOperator{\Pp}{{P}^{\prime}{}^{\mu}}
\DeclareMathOperator{\Pbar}{\bar{P}^{\mu}}
\DeclareMathOperator{\Pbarplus}{\bar{P}^{+}}
\DeclareMathOperator{\ym}{y^{--}}
\DeclareMathOperator{\Ppbra}{\langle \; P^{\prime} \; |}
\DeclareMathOperator{\Pket}{ | \; P \; \rangle}
\DeclareMathOperator{\phid}{\hat{\phi}^\dagger}
\DeclareMathOperator{\phiud}{\hat{\phi}}
\DeclareMathOperator{\dplus}{\partial^{+}}
\DeclareMathOperator{\meask}{\frac{d\kplus  d\kperp}{\sqrt{ 2 \kplus} (2\pi)^3}}
\DeclareMathOperator{\measpj}{\frac{d\pplus_{j}  d\pperp_{j}}{\sqrt{ 2 p^+_{j}} (2\pi)^3}}
\DeclareMathOperator{\measka}{\frac{d\kaplus  d\kaperp}{\sqrt{ 2 \kaplus} (2\pi)^3}}
\DeclareMathOperator{\ak}{\hat{a}(\kplus,\kperp)}
\DeclareMathOperator{\adk}{\hat{a}^\dagger(\kplus,\kperp)}

\DeclareMathOperator{\adka}{\hat{a}^\dagger(\kaplus,\kaperp)}
\DeclareMathOperator{\pket}{ | \; p_{1}, p_{2} \; \rangle}
\DeclareMathOperator{\qbra}{\langle \; p_{3}, p_{4}  \; |}

\begin{document}

\title{\hskip7cm NT@UW-01-07\\
  Exploring Skewed Parton Distributions with Two-body Models on the Light Front: bimodality}
\author{ B. C. Tiburzi and G. A. Miller\\
        Department of Physics\\ 
University of Washington\\      
Box $351560$\\  
Seattle, WA $98195-1560$ }
\date{\today}
\maketitle

\begin{abstract}
We explore skewed parton distributions for simple, model wave-functions in
a truncated two-body Fock space. Consideration of non-Gaussian 
wave functions is our main emphasis, from which we observe the distributions 
are bimodal (there can be  two distinct peaks) in the Fock-space diagonal
overlap region ($x > \xi$). We demonstrate that this behavior arises from 
convoluting two light-front wave functions to form the skewed parton distribution. 
Furthermore,  the  factorization \emph{Ans\"atze}, which have been used often,  
are shown not to hold. At high $t$, we can write the skewed parton
distribution  in terms of quark distribution amplitudes and perturbative 
wave functions thereby providing new tests. 
\end{abstract}

\section{Introduction}
Recently there has been considerable interest in the connection between hard inclusive and exclusive reactions, which has been, in part, 
due to the unifying r\^{o}le of skewed parton distributions (SPD's) \cite{Muller:1994fv, Ji:1997ek, Radyushkin:1997ki}. Aside from being the natural marriage of form
 factors and parton distributions, SPD's appear when one tries to calculate hadronic matrix elements such as in Compton scattering 
\cite{Radyushkin:1997ki, Ji:1998xh, Collins:1999be}, and the electroproduction of mesons \cite{Radyushkin:1996ru, Collins:1997fb}. 

There is much effort underway related to the measurement of
these functions \cite{Guidal:1998bk}. Intuitively clear, but simple
models \cite{Radyushkin:1997ki,Ji:1997gm,Diehl:1999kh,Choi:2001fc} have been used to provide first calculations 
(and more sophisticated approaches have been pursued \cite{Petrov:1998kf,Burkardt:2000uu}). 
In this paper, we attempt to gain intuition about the structure of these
distributions, using simple light-front wave functions. We find a new feature,
bimodality (or double peaks), which  is rather general
 and thus independent of sophisticated formalism.
 To this end, we work only with simple, two-body, model wave-functions for scalar 
particles and express the SPD as a convolution of these light-front
wave-functions.
This is, of course,
a (very) specific instance of the general light-cone Fock-space expansion of
the
SPD's carried out in \cite{Diehl:2001xz}. Of crucial importance
 is the employment of non-Gaussian wave-functions. As we shall show, the bimodal nature of SPD's is absent for Gaussian wave-functions.
 We remark that the bimodality of these distributions has been encountered before in \cite{Petrov:1998kf} where the authors study the SPD's in 
the chiral quark-soliton model of the nucleon. There it was shown that contributions from discrete levels and the Dirac continuum 
interfere, resulting in a roughly bimodal quark distribution (for $ x > \xi$). We show below that there exists a rather general explanation 
for the bimodal behavior in terms of light-front wave-functions. 

We review the definition and kinematics of the SPD in section \ref{kin} and proceed to the two-body Fock space in section \ref{FS}. 
Once in section \ref{wf}, we present the wave-functions used and then provide a visual array of SPD's illustrating their functional 
behavior. Finally we conclude briefly in section \ref{conc}.

\section{Definitions and kinematics} \label{kin}
The skewed parton distribution is a non-diagonal matrix element of bilocal field operators. Various conventions, reference frames, 
variables, \emph{etc.} exist for the description of such an object. Below we adopt the conventions of the \emph{off-forward} 
parton distribution of Ji \cite{Ji:1998pc}. This choice is natural for a symmetric treatment of initial and final hadron states and 
consequently time-reversal invariance will be manifest. 

We shall consider the SPD for a toy scalar ''meson'' of mass $M$ consisting of two scalar ''quarks'', each of mass $m$. Such a model 
restricts the SPD to only one kinematical r\'egime (one that is diagonal in Fock-space). This restriction in modeling SPD's was 
encountered from the start \cite{Diehl:1999kh},
and we take up the
issue of extending model SPD's to all kinematical r\'egimes in \cite{Tiburzi:2001je}.

Let us parameterize the kinematics as follows. The initial 
meson momentum is denoted by $\Pee = (P^{-}, P^{+}, \Pperp)$, with $ P^{\pm} = (P^{0} \pm P^{3})/\sqrt{2}$, and the
final meson momentum is denoted $\Pp$. The average momentum between initial and final states is thus
$\Pbar = (\Pee + \Pp)/2$. Defining $x \Pbarplus$ to be the momentum conjugate to 
light-front distance $\ym$, we have \cite{foot1}

\begin{equation} \label{bilocal}
F(x, \xi, t) = \frac{i \Pbarplus}{2 (\Pplus + P^{\prime}{}^{+})} \int \frac{d\ym}{2 \pi} e^{i \ym x \Pbarplus} \Ppbra \phid ( -\ym/2) 
\dplus \phiud (\ym /2) - \dplus \phid ( -\ym /2) \phiud (\ym/2)  \Pket,
\end{equation}
where the skewness $\xi$ is defined by $\xi = - \Delta^{+}/ 2 \Pbarplus$ with $\mathbf{\Delta}$ as the momentum transfer suffered 
by the meson, namely $\Delta^{\mu} = \Pp - \Pee$, with $P'_\perp=-P_\perp=\Delta_\perp/2$,
and the invariant $t \equiv \mathbf{\Delta}^{2}$. As spelled out in \cite{Ji:1998pc}, 
time-reversal invariance forces $F(x, \xi, t) = F(x, -\xi, t)$. It is easy to see why this must be the case in Eq. (\ref{bilocal}) owing 
to the reality of $F$ and the Hermiticity of the current operator. Without any loss of generality, we take $\xi > 0$ below.       
Notice the form of the above equation is such that when integrated over all $x$, we recover the electromagnetic form factor
(calculated from the $+$ component of the current operator).

Having set forth all kinematical variables, we must then use Lorentz invariance to derive any remaining relations among them. Since 
our final meson is still intact, $P^{2} = P^{\prime}{}^{2} = M^{2}$
and thus 
$\bar{P}^{2} = M^{2} - t/4$. Exploiting $\Delta \cdot \bar{P} = 0$, we find
$\Delta^{-} = \xi \bar{P}^{2}/\Pbarplus$, and also               an expression for 
the transverse momentum transfer

\begin{equation} \label{trans}
\mathbf{\Delta}^{\perp}{}^{2} = - 4 \xi^2 M^2 - t ( 1 - \xi^2).
\end{equation}
This  implies a maximal skewness $\xi_{m} = 1/\sqrt{1 - 4 M^2 /t}$ for a given $t$.

In order to unearth the physics hidden in Eq. (\ref{bilocal}),
we insert the mode expansion \cite{foot2} of the scalar field operators:

\begin{equation} \label{mode}
\phiud (\ym, \mathbf{y}^{\perp}) = \int \meask \theta(\kplus) \ak e^{-i  ( \ym \kplus - \mathbf{y}^{\perp} \cdot \kperp ) }, 
\end{equation}
with the quark creation and annihilation operators satisfying the relation

\begin{equation} \label{comm}
\big[ \ak, \adka \big] = (2 \pi)^{3} \delta(\kplus - \kaplus) \delta^{2}(\kperp - \kaperp).
\end{equation}

Evaluating the plus-derivative of the field operators and inserting this along
with the mode expansion Eq. (\ref{mode}) into the expression for the SPD, we find

\begin{equation}\label{next}
F(x, \xi, t) = \frac{1}{4} \int \meask \measka \theta(\kplus) \theta(\kaplus) \delta(x \Pbarplus - \frac{\kplus + \kaplus}{2}) 
(\kplus + \kaplus) \Ppbra \adka \ak \Pket. 
\end{equation}

\section{Two-body, truncated Fock space} \label{FS}
We now proceed with representing the SPD as a convolution of light-front wave functions. In order to do so, we must write the
meson states as sums of their light-front Fock-space components. The light-front Fock-space is the ideal arena in which to 
decompose hadronic states \cite{Brodsky:1989pv,Brodsky:2001dx}, since, apart from zero-modes, the perturbative vacuum is trivial. 
In this formalism, physical
matrix elements are then expressed as a sum of Fock-space component overlaps (in general these overlaps are non-diagonal
in particle number). The general form of the SPD written in terms of $N$-body Fock space components was carried out in 
Ref. \cite{Diehl:2001xz} and is a natural choice for these distributions since, e.g. the positivity constraints 
\cite{Pire:1999nw} are 
automatically satisfied \cite{Diehl:1999kh}. Since one has little knowledge concerning viable $N$-body light-front wavefuntions, 
this formalism
is limited to the lower Fock-space components. In order to gain initial
intuition about the SPD's, we shall limit ourselves to the two-body
sector at the cost of neglecting one kinematical region.

With a truly empty perturbative light-front vacuum $\hat{a} |0\rangle = 0$, we can build our quark states from the vacuum
in the usual fashion
\begin{equation}
|k\rangle \equiv |q; \kplus, \kperp\rangle = \adk |0\rangle.
\label{fock}\end{equation}
We shall keep the Fock-space as small as possible by truncating at quark pairs. In this approximation, our meson states appear
as
\begin{equation}\label{expand}
\Pket =  2 \Pplus (2 \pi)^{3} 
\int \Big( \prod_{j = 1}^{2} \measpj \Big) \delta^{2,+} (P - p_{1} - p_{2}) \Psi(p_{1}, p_{2}) \pket.
\end{equation} 
Several points must be clarified about the above expression. We are using shorthand notation for light-front momenta. 
Upper-case letters 
refer to meson momenta, while quark momenta are denoted by lower-case. The components $(+, \perp)$ are now implicit and we 
have notationally condensed the three-dimensional light-front delta function into $\delta^{2,+}$.

The light-front wave function $\Psi$ appears in Eq. \ref{expand}.
The remarkable property of light-front wave functions is that they 
depend only on the relative momenta of their
constituents, not on the hadron's momentum \cite{Dirac:1949cp}. Our two-body wave function is
only a function of the plus-momentum fraction $x$, and the relative transverse momentum $\pperp$, namely 

\begin{equation}
\Psi(p_{1}, p_{2}) = \psi \big(x = \frac{p_{1}^{+}}{p_{1}^{+} + p_{2}^{+}}, \pperp = (1 - x) 
\pperpone - x \pperptwo \big)
\end{equation}      
$\Psi(p_{1}, p_{2})$ is a symmetric function of $p_1$ and $p_2$. We also choose
the normalization
\begin{equation}\label{normpsi}
\int \frac{dx d\kperp}{2 (2\pi)^3 x(1-x)} | \psi(x, \kperp)|^2 = 1.
\end{equation}

It is worthwhile to state the normalization of the state $\Pket$. The use of Eqs.~(\ref{comm},\ref{fock}) gives
\begin{equation}  
\qbra \; p_{1}, p_{2} \; \rangle 
  {=} (2 \pi)^{6} \big( 
\delta^{2,+}(p_{1} - p_{3}) \delta^{2,+}(p_{2} - p_{4}). + \{ 1 \leftrightarrow 2 \} \big)
\end{equation}
Thus we have 
\begin{equation}
\Ppbra \; P \; \rangle =  2 (\Pplus)^2 \delta^{2,+}(P^{\prime} - P) \int \frac{d^{2,+} p_{1} d^{2,+} p_{2}}
{p_{1}^{+} p_{2}^{+}} \delta^{2,+}(P - p_{1} - p_{2}) \big| \Psi(p_{1},p_{2}) \big|^2  
\end{equation}
At this point, we must separate out the meson's momentum by defining total and relative momentum variables: 
$\mathcal{P} = p_{1} + p_{2}$ with $(+, \perp)$ understood, $x = p_{1}^{+}/\mathcal{P}^{+}$ and $\pperp = 
(1-x) \pperpone - x \pperptwo$. The variable differentials are related by $d^{2,+}p_{1} d^{2,+}p_{2} = 
\frac{\mathcal{P}^{+}}{2} d^{2,+}\mathcal{P} dx d\pperp$.
Carrying out the change of variables allows us to perform
the integral over $d^{2,+}\mathcal{P}$ trivially since now the delta function
$\delta^{2,+}(P - \mathcal{P})$ sits
in the integrand. The remaining integral is over relative coordinates and is merely the wave function's normalization

\begin{align} \label{norm}
\Ppbra \; P \; \rangle &=  \Pplus \delta^{2,+}(P^{\prime} - P) \int \frac{dx d\pperp}{x(1-x)} |\psi(x, \pperp)|^2 \notag \\
                       &= 2 \Pplus (2 \pi)^3  \delta^{2,+}(P^{\prime} - P).  
\end{align}

To evaluate the SPD, we insert the two-body expansion (\ref{expand}) into
equation (\ref{next})
and calculate the matrix
element
\begin{equation}
\Ppbra \adka \ak \Pket = 4 \Pplus P^{\prime}{}^{+} (2 \pi)^3 \frac{\delta^{2,+} (P^{\prime} - P + k - \kappa) 
\psi(P -k, k) \psi^{*}(\kappa, P^{\prime} - \kappa)} {\sqrt{\kplus \kaplus (P^{+} - \kplus) (P^{\prime}{}^{+} - \kaplus})}
\end{equation}
Using the delta function to perform the integral $d^{2,+}\kappa$ yields
\begin{align} \label{it}
F(x, \xi, t) &= \Pplus P^{\prime}{}^{+} \int \frac{d^{2,+}k \; ( 2 \kplus + \Delta^{+} ) \theta (x - \xi)}
{4 (2 \pi)^3 \kplus (P^{+} - \kplus)(\kplus - 2 \xi \Pbarplus)} \delta \big( (x+\xi) \Pbarplus - k^{+} \big)  
\psi(P -k, k) \psi^{*}(k + \Delta, P - k) \notag \\
&= \frac{x (1-\xi^2) \; \theta ( x - \xi) }{2 (2 \pi)^3 (1-x) (x^2 - \xi^2)} 
\int d\kperp \psi^{*}\Big(\frac{x- \xi}{1- \xi} ,\kperp + \frac{1-x}{1-\xi^2} 
\mathbf{\Delta}^{\perp} \Big) \psi \Big( \frac{x+\xi}{1+\xi},\kperp \Big),  
\end{align}
where we see the result of the theta functions present in Eq.~(\ref{next})
is to restrict the value of $x$: $1 > x > \xi$. In the other kinematical r\'egime,  $-\xi < x <  \xi$, the SPD 
represents the amplitude to 
annihilate a quark pair from the initial meson state. Naturally the SPD in this r\'egime depends upon the higher Fock components,
which we neglect here.

The above result is consistent with the time-reversal invariance of our
starting expression Eq. (\ref{bilocal}) since 
it is clearly
even in $\xi$. In the forward limit $(t = 0)$, we recover the toy model quark distribution function
\begin{equation} \label{toyq}
F(x, 0, 0) = \int \frac{d\kperp}{2 (2\pi)^3 x (1-x)} |\psi(x, \kperp)|^2 \equiv q(x).
\end{equation}
Finally in the case of zero skewness we should arrive at the electromagnetic form-factor (after having integrated over
$x$) \cite{Radyushkin:1997ki}. We find
\begin{equation} \label{subs}
\int dx F(x,0,t) = \int \frac{dx d\kperp}{2 (2\pi)^3 x (1-x)} \psi(x, \kperp) \psi^{*}(x, \kperp + (1-x) \mathbf{\Delta}^{\perp})
\equiv F(t), 
\end{equation}
which is indeed the Drell-Yan formula \cite{Drell:1970km}. As pointed out in
Ref. \cite{Diehl:1999kh},
the  restriction to one kinematical
r\'egime ($x > \xi$) does not allow for the sum rule to be
verified---Eq. (\ref{subs})
is as much as we can hope for now. We shall take up the issue of verifying the sum rule in Ref.~\cite{Tiburzi:2001je}.

Note that the form of the SPD does not support various factorization \emph{Ans\"atze}\cite{Vanderhaeghen:1998uc}.  For example, 
in the limit $\xi \to 0$,  $F(x, \xi, t) \ne q(x) F(t)$, though this \emph{Ansatz} does satisfy the sum rule. 
Such a factorized form only holds for Gaussian wave functions when $\xi = 0$. Counting the number of field operators 
in equation \ref{bilocal}, we see there is no physical basis for such a
factorized \emph{Ansatz} when $\xi=0$.    

\section{Model wave functions} \label{wf}
It remains only to choose two-body light-front wave functions to explore the
SPD derived in
Eq. (\ref{it}). Before we do so, it is wise 
to review the existing literature on modeling SPD's. Radyushkin investigated them for a scalar toy model \cite{Radyushkin:1997ki} deriving analytical 
results. They were investigated for a bag model of the nucleon in \cite{Ji:1997gm}, in the chiral quark-soliton model
\cite{Petrov:1998kf} and in $1+1$ dimensional QCD \cite{Burkardt:2000uu}.
More in line with our above analysis, SPD's were modeled using light-front
 Fock-space components in \cite{Diehl:1999kh} (using $N$-body Gaussian wave functions
 for low values of $N$),
 for the pion using a Gaussian wave function in\cite{Choi:2001fc}, and for the QED 
electron in \cite{Brodsky:2001xy}. 

We will use three different wave functions to contrast the behavior of the SPD: a Gaussian (G), the Hulth\'en wave function (H) of
 Ref. \cite{Tiburzi:2000je}, and the (weak-binding) Wick-Cutkosky wave function (WC) of Ref. \cite{Karmanov:1980if}. Each wave function contains a 
normalization constant $N$ (determined from the normalization
condition used
above in Eq. (\ref{normpsi})) and is chosen to be symmetric about $x = 1/2$.

Our Gaussian wave function is taken to be
\begin{equation}
\psi_{G}(x, \kperp) = \frac{\sqrt{N f(x)}}{m} e^{-\frac{\kperp^2 + m^2}
{\mu^2 x (1-x)}},
\end{equation}
where $f(x) = x^{2} (1-x)^{2}$
is similar to a valence quark distribution (the difference from the quark 
distribution of Eq.(\ref{toyq}) is due to the mass term in the exponential). We choose $\frac{m}{\mu} = 1$ and $\frac{M}{\mu} = 1.8$. 

The Hulth\'en wave function has the form
\begin{equation}
\psi_{H}(x, \kperp) = \frac{m \sqrt{N} x (1-x)}{4 x (1-x) \alpha^2 + (2 x - 1)^2 m^2 + \kperp^2} 
\big( \delta_{a}^{\alpha} - \delta_{b}^{\alpha} \big), 
\end{equation}
with $\frac{a}{m},\frac{b}{m}$ chosen to be the phenomenological deuteron parameters. 

Lastly the Wick-Cutkosky wave function in the weak-binding limit is
\begin{equation}
\psi_{WC}(x, \kperp) = \frac{m^3 \sqrt{N} x^2 (1-x)^2}{(4 x(1-x) \kappa^2 + (2 x -1)^2 m^2 + \kperp^2)^2} \frac{1}{1 + |2 x - 1|},
\end{equation}
with $\kappa = \frac{1}{2} m a$, the meson mass $M = 2m - \frac{1}{4} m a^2$. We choose a suitably weak coupling of $a = 0.08$. The
form of the wave function is basically Coulombic with a multiplicative retardation term. 

\begin{figure}
\begin{center}
\epsfig{file=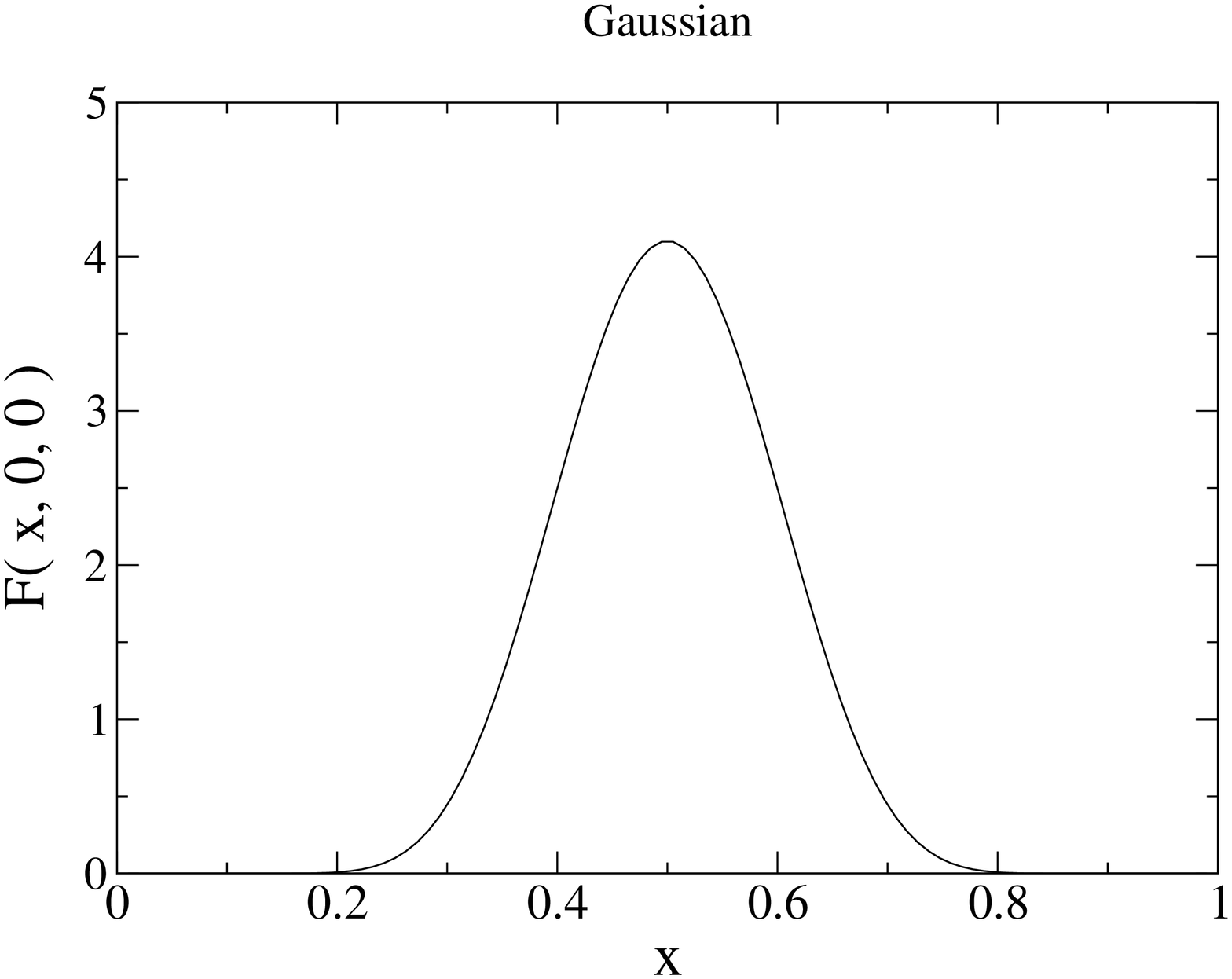,height=3.5in,width=1.5in,angle=270}
\epsfig{file=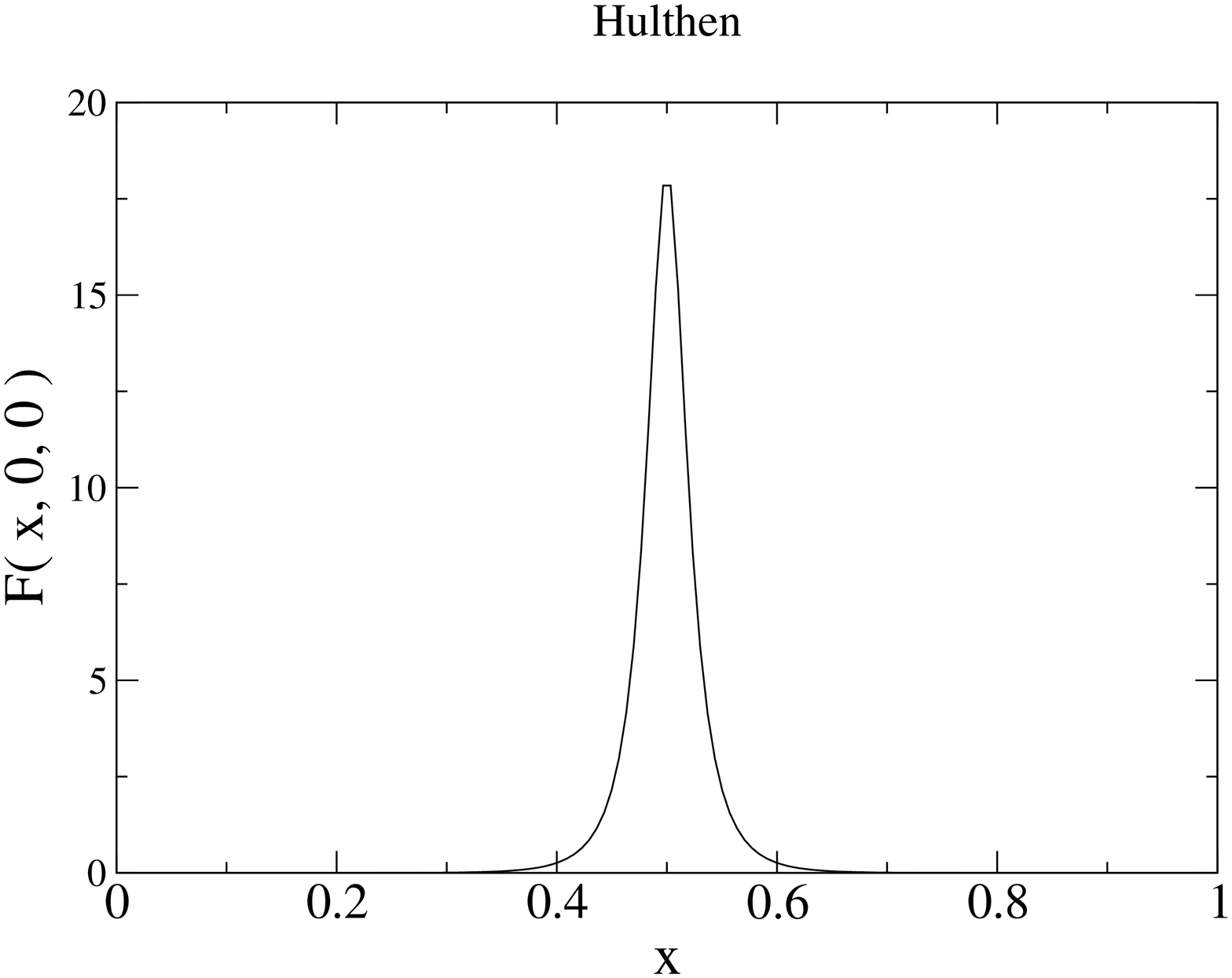,height=3.5in,width=1.5in,angle=270}
\epsfig{file=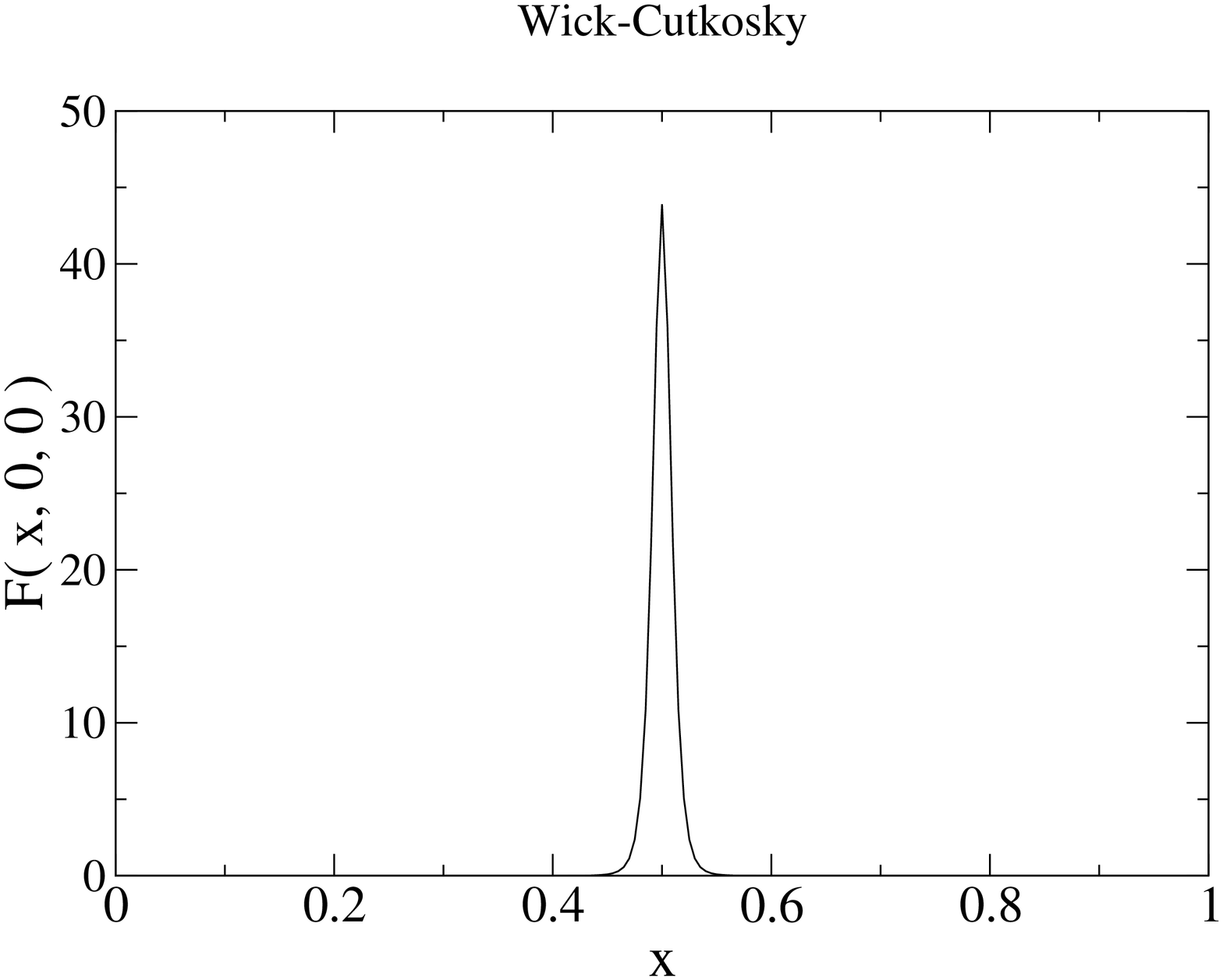,height=3.5in,width=1.5in,angle=270}
\caption{Quark distribution functions $F(x, 0, 0)$ plotted verses $x$ for the models under consideration.}
\label{fdist}
\end{center}
\end{figure}

In Figure \ref{fdist} we contrast the quark distribution function for each of the
toy models under consideration. The models clearly differ in how narrowly the 
quark distribution is peaked. The Wick-Cutkosky distribution appears sharply 
peaked at $x = 1/2$ which is characteristic of a weakly bound system. 

\begin{figure}
\begin{center}
\epsfig{file=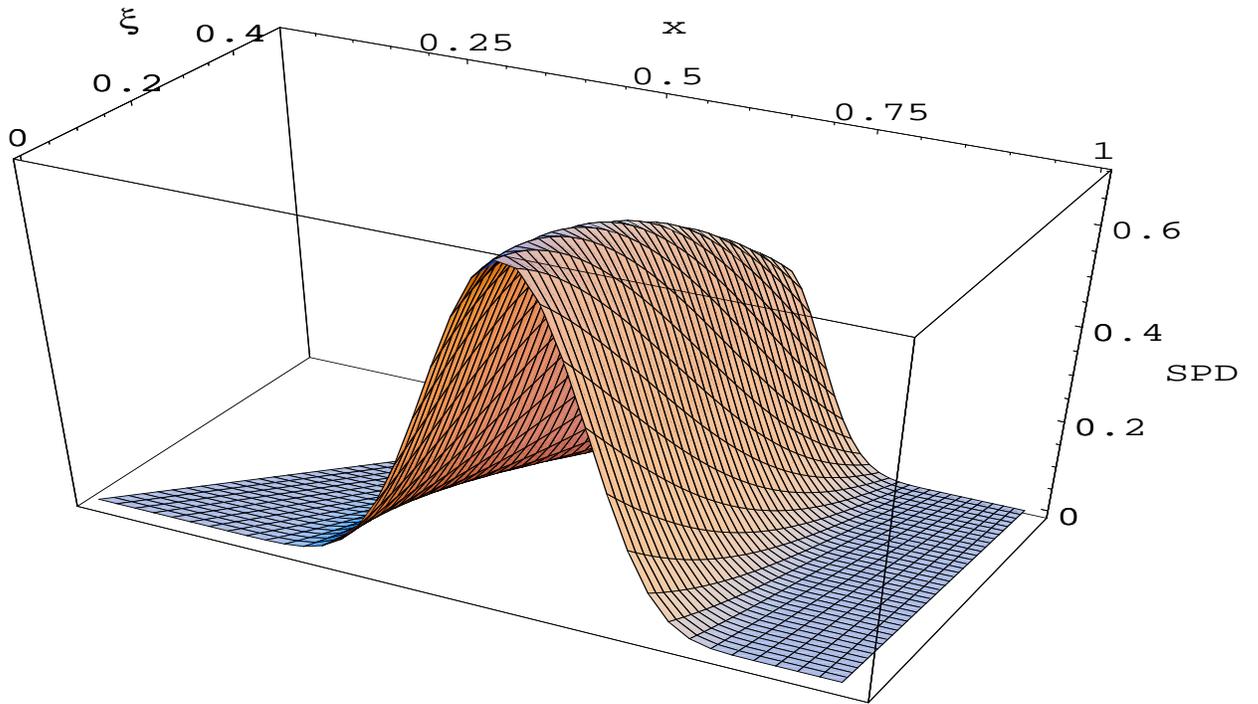,height=4in,width=6.5in}
\caption{Skewed parton distribution for 2-body Gaussian model using the 
parameters: $m/\mu = 1, M/\mu = 1.8$ and evaluated for $-t/M^2 = 1.23$.}
\label{fgauss}
\end{center}
\end{figure}

Next in Figure \ref{fgauss}, we use our Gaussian wave function $\psi_{G}$ in
Eq.~(\ref{it})
to reproduce a typical Gaussian SPD 
found in the literature. We note the distribution maintains its characteristic Gaussian shape as a function of $\xi$. 
The peak of the distribution, however, appears to be suppressed as $\xi$ increases up to $\xi_{m}$. Since (as we will show) 
the special form of the Gaussian wave function obscures the general structure of the SPD, we proceed to consider the power-law 
wave functions.

\begin{figure}
\begin{center}
\epsfig{file=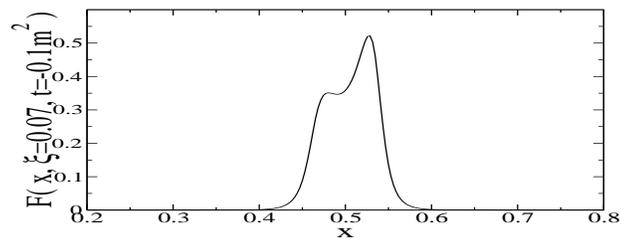, height=3.5in, width=1.5in,angle=270}
\caption{Bimodality encountered in the Wick-Cutkosky Model's skewed 
parton distribution. The SPD is plotted as a function of $x$ for $-t / m^2 = 
0.1$ and $\xi = 0.07$.}
\label{fwc1}
\end{center}
\end{figure}

\begin{figure}
\begin{center}
\epsfig{file=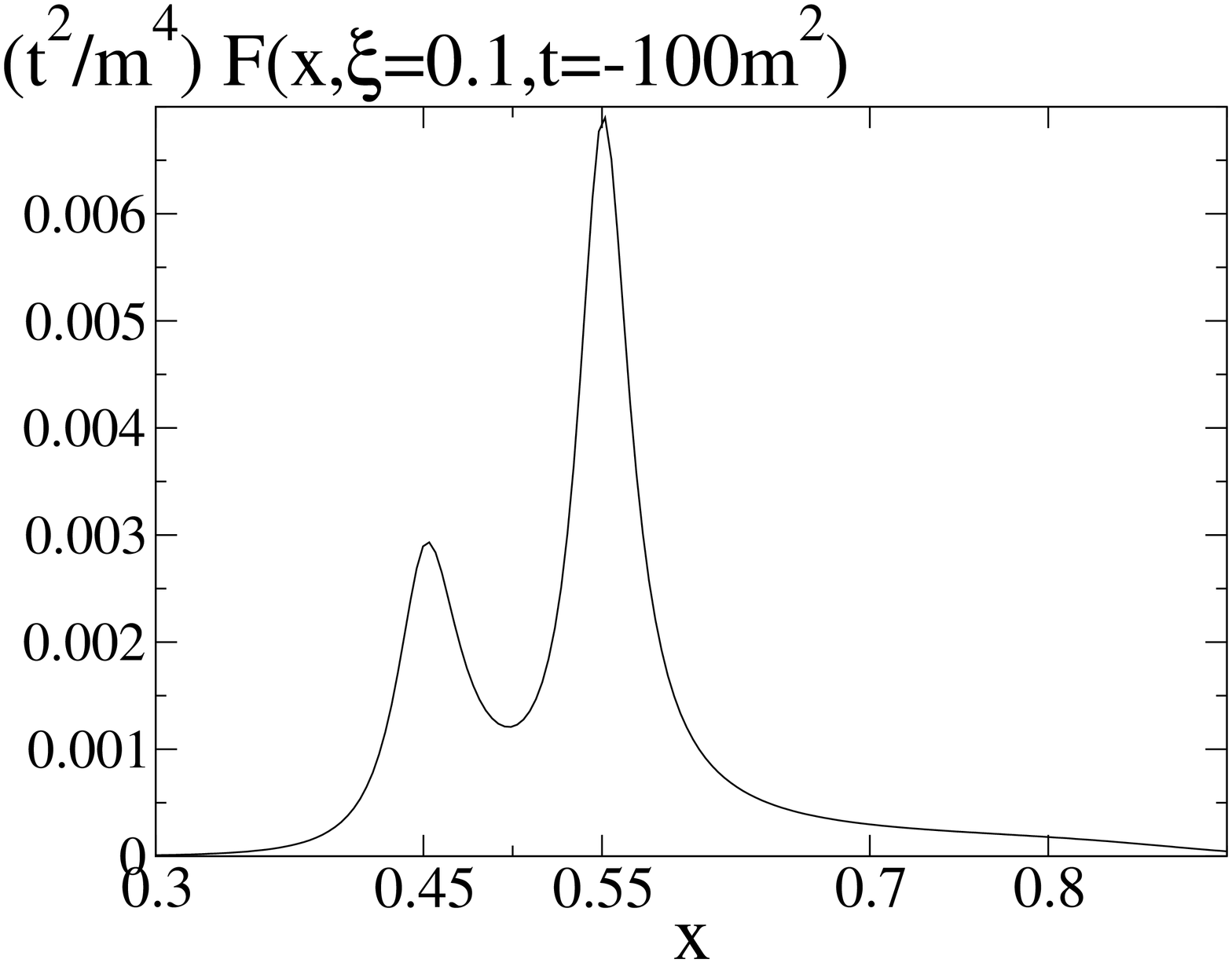, height=3.0in, width=1.5in,angle=270}
\caption{Bimodality in the Wick-Cutkosky Model's skewed
parton distribution at high momentum transfer. The SPD is plotted as a function of $x$ for $-t / m^2 = 
100$ and $\xi = 0.1$.}
\label{fwc2}
\end{center}
\end{figure}

Let us plot the SPD using Eq.~(\ref{it})
for the Wick-Cutkosky wave function $\psi_{WC}$. In Figure \ref{fwc1}, we show the $\xi = 0.07$ 
slice of the SPD at $-t/m^2 = 0.1$. The distribution is bimodal, with the two peaks displaced from $x = 1/2$. To understand this
feature, we crank up the momentum transfer to $-t/m^2 = 100$. The larger momentum transfer allows us to probe a larger skewness (since
$\xi_{m}$ increases with $-t$). At this higher momentum transfer, let us look at another $\xi$-slice of the SPD as a function of $x$ 
shown in Figure \ref{fwc2}. 

\begin{figure}
\begin{center}
\epsfig{file=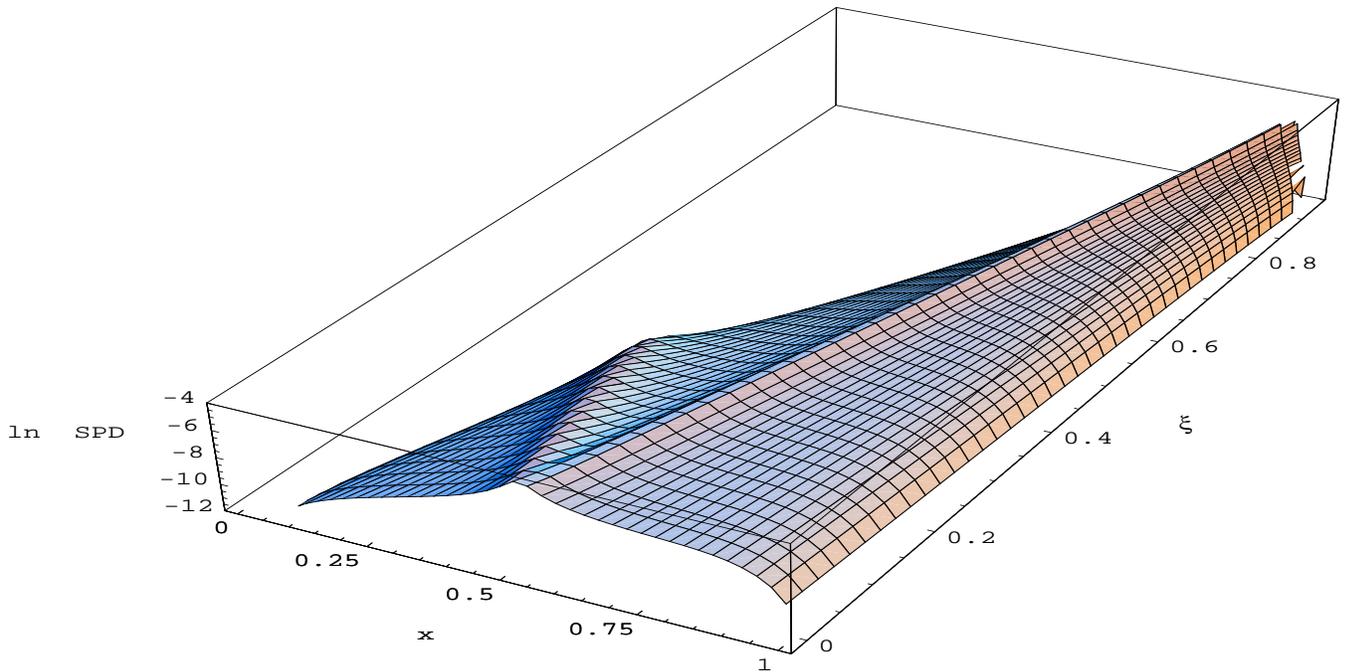,height=4in,width=7in}
\caption{Skewed parton distribution for Wick-Cutkosky Model, with the model parameter $a = 0.08$, evaluated for $-t/m^2 = 100$. 
This asymptotic momentum transfer allows us to see and understand the general features of the distribution. Notice we plot the 
natural logarithm of the SPD.}
\label{fwc3}
\end{center}
\end{figure}

Perhaps the bimodal nature of the distribution is not too surprising, after all the SPD is a convolution of two hadronic wave 
functions. The weak binding Wick-Cutkosky wave function is by nature highly peaked at $x = 1/2$. The integration over the 
transverse momentum $\kperp$ in Eq.~(\ref{it})
does not affect the location of the peak. Due to the convoluted nature of the SPD, 
the $x$ location of
the wave function's peak becomes a function of $\xi$. Let us denote
the location of $\psi$'s peak---$\psi(\;\alpha(x,\xi), \kperp)$ with the arguments in Eq.~(\ref{it})---as $x_{p}(\xi)$ 
and that of $\psi^{*}$ by $\bar{x}_{p}(\xi)$. Using Eq.~(\ref{it}), we find

\begin{align}
x_{p}(\xi)       & = \frac{1}{2} ( 1 - \xi) \notag \\
\bar{x}_{p}(\xi) & = \frac{1}{2} ( 1 + \xi).\notag
\end{align}
Thus for fixed $t$, as the skewness $\xi$ is increased, the peaks of $\psi$ and $\psi^{*}$ trek in opposite directions. For 
$\xi =  0.1$, the peak of $\psi$ should be located at $x_{p} = 0.45$ and that of $\psi^{*}$ at $\bar{x}_{p} = 0.55$. Looking back
at Figure \ref{fwc2} shows us this is indeed the case. This allows us to understand the peaks' behavior for the full range of $\xi$
shown in Figure \ref{fwc3}, where we see a ripple which heads to smaller $x$ as $\xi$ increases. Once $\xi > 1/3$ we can be 
assured the SPD's peak is due to $\psi^{*}$. But the figure opens up new questions. What can we say about the height of the 
peaks?

Looking at all the SPD's displayed so far for the Wick-Cutkosky model, generally the peaks at lower $x$
which stem from $\psi$ aren't quite the lofty summits as those from $\psi^{*}$ which are at higher $x$. 
This asymmetry is due to the prefactor in the SPD: $\frac{x(1-\xi^2)}{(1-x)(x^2 -\xi^2)}$ which
generally increases with increasing $x$ for fixed values of $\xi$. 
Thus the peaks due to $\psi^{*}$ are preferentially 
enhanced while those due to $\psi$ tend to be flattened as they trek to lower $x$. 

Taking a second glance at Figure \ref{fwc3}, we see another feature of these distributions. The peak which we attribute to 
$\psi^{*}$ is amplified with increasing $\xi$. To be able to make
definite statements,  we take  the asymptotic limit and 
appropriately analyze the SPD.
(We will find that this leads to generally true qualitative information for lower momentum transfer.) 
The Wick-Cutkosky wave function has power-law dependence, and so we appeal to the
approach of Brodsky and Lepage \cite{Brodsky:1989pv}
of approximating  integrals over ${\bf  k}_\perp$  using the knowledge that our
wave functions are peaked for low values of $\kperp$. Looking at Eq.~(\ref{it}) and
taking $\Delta^\perp$ very large compared to  ${\bf  k}_\perp$, we realize
the dominant contributions come from either ${\bf  k}_\perp\approx0$ or
${\bf  k}_\perp \approx - \frac{1-x}{ 1-\xi^2} \Delta_\perp$ for which one of the
wave functions is greatly suppressed. This gives:
\begin{equation} \label{asymf}
F(x, \xi, t/m^2 \gg 1) \approx  \frac{x(1-\xi^2)}{2 (2\pi)^3 (1-x) (x^2 - \xi^2)} \Bigg(  \phi (x_{1}) \psi^{*}(x_{2},(1-\bar{x})
\mathbf{\Delta}^{\perp} ) + \phi^{*}(x_{2}) \psi(x_{1}, -(1-\bar{x}) \mathbf{\Delta}^{\perp}) \Bigg),
\end{equation}
where $\phi(x) = \int d\kperp \psi(x,\kperp)$ and we've abbreviated the plus-momentum fraction arguments as 
$x_{1} = \frac{x+ \xi}{1+ \xi}$, $x_{2} = \frac{x - \xi}{1 - \xi}$ and their
average $\bar{x} = (x_{1} + x_{2}) /2$. Notice that the   
above equation reproduces the correct 
location of the peaks as a function of $x$. The Wick-Cutkosky wave function, as well as the Hulth\'en
behave as $\psi \sim 1/t^2$ for large $|t|$. In a realistic calculation, one would use wave functions 
calculated from perturbative Quantum Chromodynamics (pQCD). These wave functions have different asymptotic behavior
($\psi_{\text{pQCD}} \sim 1/t$) than our model wave functions since the dynamics of pQCD stems from vector exchange
(unlike the scalar exchange implicitly employed by our models). The pQCD analysis of the pion's SPD was carried out
in \cite{Vogt:2001if}.

Using the form of Eq.~(\ref{asymf}), we can deduce whether the wave functions are being stressed by increasing $\xi$. 
The wave functions will be stressed depending on whether larger transverse momentum flows through them as $\xi$ increases. 
Thus we are concerned as to whether $(1-\bar{x}) \Delta^{\perp}$ increases with $\xi$ (for fixed $t$ and $x$). Taking the 
derivative with respect to $\xi$
shows that the change in transversal momentum flow
depends upon the sign of $|t| ( 1- \xi^2 ) - ( 1 + \xi^2) 4 M^2$. 
Hence, if $|t|/4 M^2 < 1$, then less momentum will flow through
the wave functions as $\xi$ increases. Otherwise, there is initially
increasing momentum flow (starting from $\xi = 0$)
followed by decreasing momentum flow with $\xi$ \cite{infl}. In either case, 
as $\xi$ approaches its maximum value (the maximal skewness $\xi_{m}$),
there is less transverse momentum flowing through the wave functions in Eq.~(\ref{asymf}) and 
consequently the height of the summit (centered at $\bar{x}_{p}$) will increase. 

\begin{figure}
\begin{center}
\epsfig{file=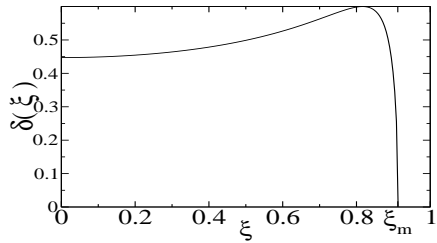, height=2.5in,width=1.5in,angle=270}
\caption{Ratio of transverse momentum flow to $M$: $\delta(\xi) = (1 - \bar{x}) |\Delta^{\perp}|/M$, plotted as
a function of $\xi$ (taken here for $t/M = - 20$  and $x = 0.9$). Note the drastic drop as we 
approach maximal skewness.}
\label{f5a}
\end{center}
\end{figure}

This fact is evidenced by Figure \ref{fwc3}. Given the (natural) logarithmic scale, we see the SPD increases by a factor $\sim 10$ 
near maximal skewness. To quantitatively understand this rise in the distribution, we'd be wise to consider the 
transverse momentum flowing through the wave functions. The ratio of the transverse momentum flow (in asymptopia)
to the mass is given by  $\delta(\xi) = (1 - \bar{x}) |\Delta^{\perp}|/M$. For fixed $x$ (say near the peak of $\psi^{*}$ as we
go to maximal skewness) and fixed $t$, we can plot this ratio as a function of $\xi$ which is done in Figure
\ref{f5a}. Indeed near maximal skewness, there is a drop ($\sim 10 \times$) in $\delta$ and consequently we
expect a rapid rise in the distribution since (rather suddenly) the wave functions aren't as stressed.

\begin{figure}
\begin{center}
\epsfig{file=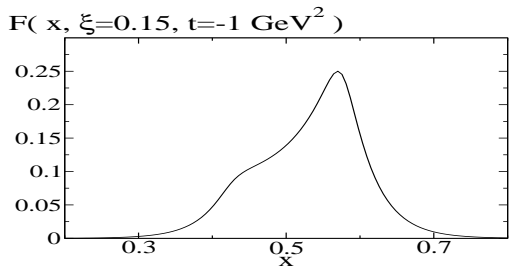,height=3.0in,width=1.5in,angle=270}
\caption{Hulth\'en SPD for $|t| = 1$ GeV${}^2$ and $\xi = 0.15$ illustrating the bimodality.} 
\label{fh1}
\end{center}
\end{figure}

\begin{figure}
\begin{center}
\epsfig{file=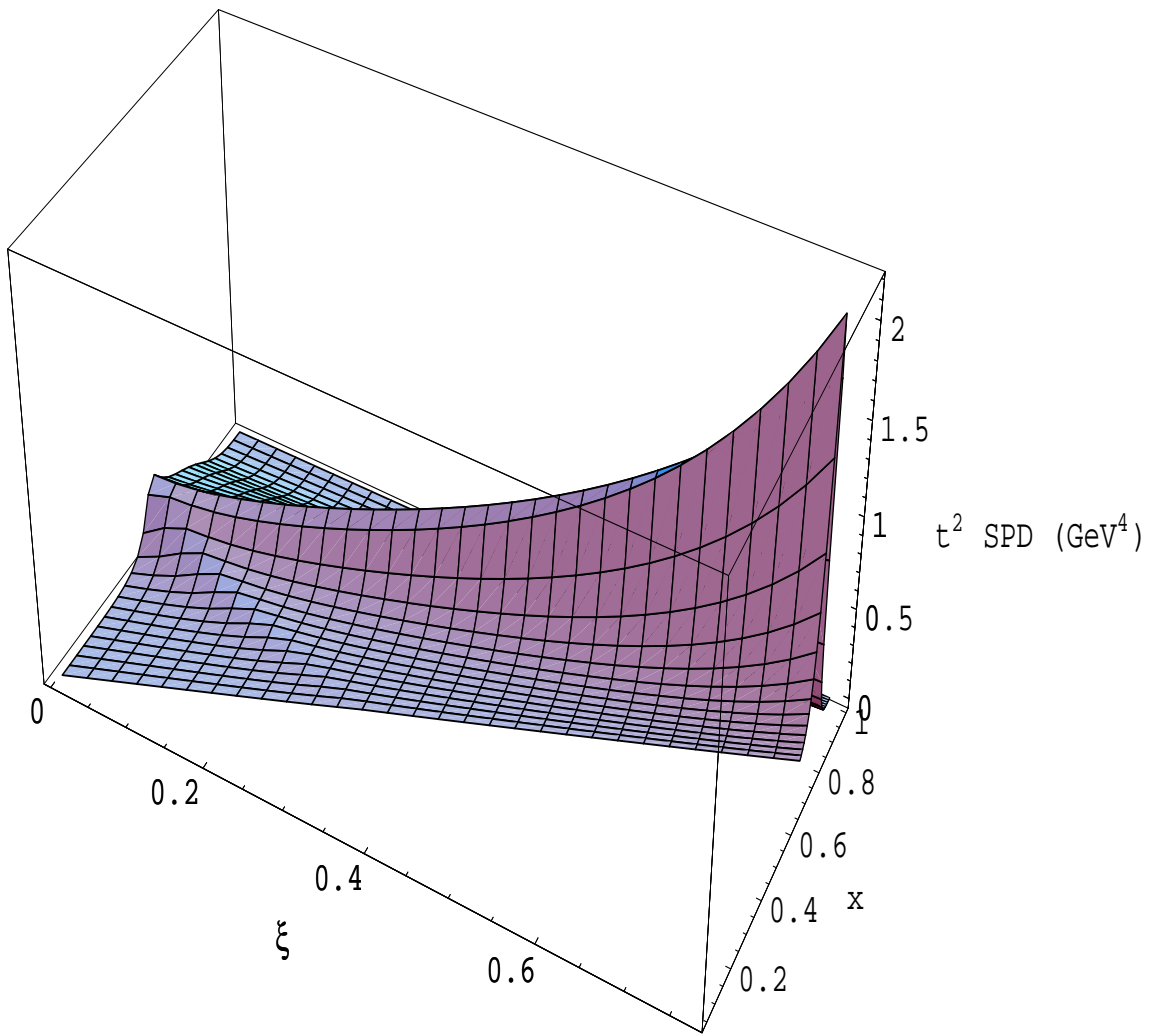,height=7in,width=6in}
\caption{Skewed  parton distribution for the Hulth\'en model plotted for $|t| = 20$ GeV${}^2$. The trends at this high momentum
transfer are now identical to those in the Wick-Cutkosky model.}
\label{fh2}
\end{center}
\end{figure}

The arguments presented above have been general for power-law wave functions and so aptly apply to the Hulth\'en model 
wave function as well. We can find the SPD's for the Hulth\'en model by
inserting $\psi_{H}$ into equation \ref{it}.
At the moderate
momentum transfer of $1$ GeV${}^2$
the bimodality is visible in the $\xi = 0.15$ slice of the SPD, see Figure \ref{fh1}.
The peaks are not quite so distinct since the Hulth\'en wave function isn't as narrowly peaked about $x = 1/2$ compared to the 
Wick-Cutkosky. Consequently distinguishability between contributions from $\psi$ and $\psi^{*}$ diminishes. As we go to higher 
momentum transfer, the trends become as clear as before. Figure \ref{fh2} plots the SPD at $|t| = 20$ GeV${}^2$ which matches 
up to the Wick-Cutkosky SPD (Figure \ref{fwc3}) in every detail. At $20$ GeV${}^2$, the maximal skewness is $\xi_{m} \approx 0.77$.
In the plot, we have shown the full range of $\xi$.  The bimodality is still visible for $\xi < 1/3$ but the scale is
dominated by the rise toward maximal skewness. We ran into to this in the Wick-Cutkosky model (and was the reason we previously 
used the logarithmic scale in Figure \ref{fwc3}). Taking our focus away from the bimodal peaks, we see the distribution
rises ($\sim 5\times$) as $\xi$ approaches maximal skewness even against the competing prefactor $1-\xi^2$. The power-law 
Hulth\'en wave function is clearly recovering from being stressed as we approach maximal skewness. 

\begin{figure}
\begin{center}
\epsfig{file=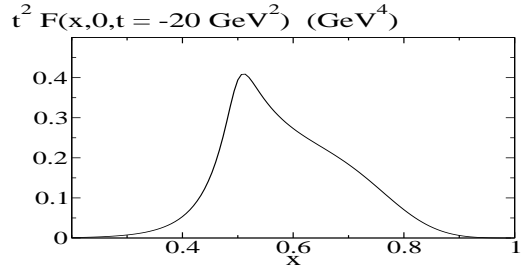,height=3in,width=1.5in,angle=270}
\caption{Un-$x$-integrated Hulth\'en form factor for $|t| = 20$ GeV${}^2$, i.e. the SPD evaluated for $\xi = 0$.
The factorization violating peak has already started to develop (manifesting itself as an asymmetry).}
\label{fh3}
\end{center}
\end{figure}

Finally we must remark, the altered view of Figure \ref{fh2} not only allows us to view
the bimodality more clearly, it also hides a feature of in our model wave function! 
If we could look into the large $x$ shadows in Figure \ref{fh2}, we would 
notice a striking asymmetry present even at $\xi = 0$. The cross section shown in 
Figure \ref{fh3} highlights this feature. This built-in asymmetry persists in the SPD for 
$\xi \ne 0$ \cite{well2}. Recall 
that the SPD at $\xi = 0$
gives the un-$x$-integrated electromagnetic form factor. In light of our previous work with
the Hulth\'en model form factor \cite{Tiburzi:2000je},
we readily interpret the asymmetry as due to factorization breaking present in this model.
The dominant contribution to the electromagnetic form factor in asymptopia comes from a peak in the near end-point region ($x \lesssim 1$)
and dominates the contribution from the peak at $x = 1/2$ by a factor of $\ln |t|$. As $|t|$ is increased the asymmetry shown in Figure
\ref{fh3} will develop into a strong peak in the near end-point region giving rise to the logarithmic modification to the form
factor. 

Of course logarithmic, asymptotic  modifications to the form factor are also present for the Wick-Cutkosky model 
\cite{Karmanov:1992fv}. 
One finds a near end-point asymmetry for the Wick-Cutkosky SPD as well, however, due to the value of model
parameter chosen ($a = 0.08$), one must go to considerably high
momentum transfer ($|t|/m^2 \sim 10^5$) to see this effect.

Having visited the power-law, model wave functions, we must come full circle.
What about the Gaussian wave function we initially 
considered? We must remark that the logic leading to Eq.~(\ref{asymf})
depends crucially on the fact that the wave functions have 
power-law momentum dependence. Thus arguments
about transversely stressed wave functions simply do not apply and we cannot gain
intuition about the SPD's amplitude.
Moreover (owing to the peculiar nature of Gaussians) taking the convolution of two Gaussian wave 
functions yields another Gaussian wave function (certainly with a different \emph{effective} plus-momentum, but a Gaussian nonetheless).
Thus bimodality (which was due to the individual nature of $\psi$ and $\psi^{*}$) must disappear. Moreover, at large momentum transfer, 
we do not see an amplification of the SPD near maximal skewness. On the contrary, the Gaussian SPD's are suppressed.
 Indeed the behavior of Gaussian wave 
functions is peculiar (not just due to unimodality). Luckily we have a wealth of intuition about power-law wave functions and thus 
our understanding of the structure of skewed parton distributions need not remain cloudy. The amplification (with $\xi$) of the 
SPD is a possible signature for power-law dependence. Physical observables (the deeply virtual Compton scattering 
cross section, for example) depend on weighted $x$-integrals of the SPD. Thus maximal skewness amplification could be
canceled by compensating behavior in the other kinematical r\'egimes. 

\section{Concluding remarks} \label{conc}
Above we have derived the form of the skewed parton distribution in terms of two-body light-front wave functions in the region
$x > \xi$. We then used model wave functions in order to explore the resulting distributions and gain intuition about their structure. 
The key results are: Eq.~(\ref{it}), that the use  of power-law wave functions leads to skewed parton distributions with two peaks, 
that factorization of $F(x,\xi=0,t)$ of  into separate functions of $x$ and $t$ does not hold, and finally Eq.(\ref{asymf}). 
This last item gives  $F(x,\xi,t)$ for very large values of $t$ in terms of quark distribution amplitudes and perturbative wave functions. 
Thus SPD's give a new way to investigate perturbative treatments.

\begin{center}
{\bf Acknowledgment}
\end{center}
This work was funded by the U.~S.~Department of Energy, grant: DE-FG$03-97$ER$41014$.


\begin{thebibliography}{99}
\bibitem{foot1}
Our expression for the SPD tacitly employs the light-front gauge in which $A^{+}=0$.
\bibitem{foot2}
 Since our Fock-space expansion will be truncated to just quark
pairs, the results
will not depend on whether we retain anti-particles in the mode expansion.
This is done to  eliminate from the start kinematical r\'{e}gimes not under consideration.
\bibitem{infl}
One can show that where the derivative vanishes ($\xi_{0}$) always occurs
before
the maximal skewness, \emph{vis.} $\xi_{0} < \xi_{m}$. 
Thus one is always guaranteed that less transversal momentum will flow through the wave functions for \emph{physical} $\xi$ which 
are large enough.
\end{thebibliography}
\end{document}